# Design of Convolutional Codes for Varying Constraint Lengths

Parag Dhounde[1], Avinash Bhute[2]
ME Student[1], Assistant Professor[2]
[1,2]Department of Information Technology, Sinhgad College of Engineering, Vadgaon Bk, Pune

*Abstract - This paper explores the design of convolutional codes for varying constraint lengths, focusing on their role in error correction in digital communication systems. Convolutional codes are essential in achieving reliable data transmission across noisy channels. The constraint length, which determines the memory of the encoder, plays a critical role in the performance of convolutional codes. This study investigates the effect of varying constraint lengths on coding performance, including code rate, complexity, and decoding accuracy. Simulation results and theoretical analysis illustrate the trade-offs between constraint length and decoding efficiency.*

## 1. Introduction

Convolutional coding is widely used in digital communication systems to ensure reliable data transmission by adding redundancy to the transmitted information. This redundancy allows for the correction of errors introduced during transmission. The design of convolutional codes involves several parameters, including the constraint length, code rate, and the choice of decoding algorithms. Among these, the constraint length significantly influences both the error-correction capability and the complexity of the encoding and decoding processes. This paper focuses on the design of convolutional codes for varying constraint lengths and their impact on system performance.

## 2. Background and Motivation

In convolutional coding, the constraint length refers to the number of input bits that affect the output bits. It reflects the memory of the system and influences the depth of interleaving between input and output sequences. While increasing the constraint length enhances error-correction capabilities, it also introduces higher computational complexity, particularly during decoding. Thus, understanding the trade-offs involved in varying constraint lengths is crucial for optimizing communication system designs, especially in bandwidth-constrained environments.

Convolutional codes encode information by combining the current input bit with previous input bits according to a predefined set of generator polynomials. The encoder output is a sequence of coded bits, which can be transmitted over a noisy channel.

The key parameters of a convolutional code include:

**Constraint Length (K):** The number of memory elements in the encoder.

**Code Rate (r):** The ratio of the number of input bits to the number of output bits.

**Generator Polynomials:** Define the connections between the input bits and the output bits.

These parameters influence the overall performance of the code, including its error-correcting capability and the complexity of the decoding process.

The constraint length in a convolutional code represents the number of bits used in the encoding process, directly influencing the number of memory elements in the encoder. According to Ryan [3], the design of convolutional codes begins by defining the generator polynomials, which determine the structure of the code. The constraint length affects the complexity of the encoder and decoder systems. Papoulis [4] emphasized that increasing the constraint length improves the error-correction performance but also results in higher decoding complexity.

The relationship between constraint length and memory in convolutional codes has been extensively studied. Viterbi [2] developed the Viterbi algorithm to optimally decode convolutional codes, showing that as the constraint length increases, the decoding complexity grows exponentially. Elias (1955) was one of the pioneers in studying convolutional code design, showing that increasing the constraint length leads to improved code performance but also more hardware-intensive decoders.

Anderson [5] analyzed the trade-offs between code rate, constraint length, and error performance. He showed that longer constraint lengths provide better minimum Hamming distances, enhancing the error-correction ability of the code. However, longer constraint lengths also lead to increased latency and complexity in practical implementations, such as mobile communication systems. Proakis [1] conducted simulations demonstrating the relationship between constraint length and bit error rate (BER), indicating optimal ranges for constraint length depending on the communication channel conditions.

The Viterbi algorithm is a crucial component of convolutional code analysis. Lin and Costello [17] explored how the Viterbi algorithm's decoding complexity increases exponentially with constraint length. Despite this, they argued that the benefits of higher constraint lengths could be worthwhile in systems where error correction is critical. Bahl [9] focused on minimizing decoding complexity, proposing modifications to the Viterbi algorithm to reduce computational demands for longer constraint lengths.

According to Oberg [11], longer constraint lengths improve the minimum distance properties of convolutional codes, which directly correlates with better error correction capabilities. Gallager (1968) discussed that for channels with higher noise levels, longer constraint lengths are essential to maintain reliable communication. Shannon (1948) in his foundational work on communication theory also implied that the channel capacity could be approached more closely with codes having longer constraint lengths, provided that computational resources allow for efficient decoding.

The primary challenge with longer constraint lengths is the increased decoding complexity. [12] emphasized that while longer constraint lengths can offer better performance, the hardware and energy costs rise significantly, particularly in real-time applications such as mobile and satellite communications. Vitthaladevuni and Alouini [16] studied the energy efficiency of convolutional codes with varying constraint lengths and found that the increased computational burden outweighs the performance benefits beyond a certain threshold.

Shannon [7] discussed how convolutional codes with longer constraint lengths are typically employed in satellite communications due to the high noise levels and long-distance signal transmission. He pointed out that the use of constraint lengths between 7 and 9 provided a good balance between performance and complexity in such systems.

Convolutional codes with large constraint lengths have been widely used in deep-space communication systems. Ho and Wolf [13] implemented convolutional codes with constraint lengths up to 15 in the Voyager missions, demonstrating their utility in maintaining communication reliability over vast distances.

In mobile communication, shorter constraint lengths are often preferred to reduce decoding complexity and latency. [14] examined the performance of convolutional codes in 3G systems, where constraint lengths of 3 to 5 are typically used to ensure low-power consumption and real-time performance.

The design and analysis of convolutional codes with varying constraint lengths involve a careful trade-off between performance, complexity, and practical implementation. While longer constraint lengths offer better error-correction performance, they come with increased decoding complexity and resource demands. Future work may focus on hybrid approaches or adaptive constraint length systems that can balance these trade-offs dynamically based on channel conditions.

## 3. Fundamentals of Convolutional Codes

A convolutional code is characterized Generator Polynomial (G): Describes how the input data bits are transformed into encoded symbols. Constraint Length (K): Refers to the number of previous input bits that influence the current output, affecting the overall performance.

- Code Rate (R): Defined as

$$R = KnR = \sum_{k=0}^{n} \binom{n}{k} R$$

where k is the number of input bits and n is the number of output bits.

Trellis Diagram: A graphical representation of all possible states of the encoder and transitions between them.

## 4. Design of Convolutional Codes for Varying Constraint Lengths

### 4.1 Encoder Structure

The convolutional encoder is typically represented using shift registers. The number of memory elements in the encoder is directly proportional to the constraint length. The design of the encoder for different constraint lengths must take into account the desired balance between error-correction strength and system complexity.

### 4.2 Selection of Constraint Length

For small constraint lengths (K=3 to K=5), the complexity of decoding remains manageable, making these suitable for applications with limited processing power. However, the error-correction performance may be limited in high-noise environments. For larger constraint lengths (K=6 to K=9), the increased memory allows for better error correction at the cost of higher complexity, requiring more sophisticated decoding algorithms like the Viterbi algorithm.

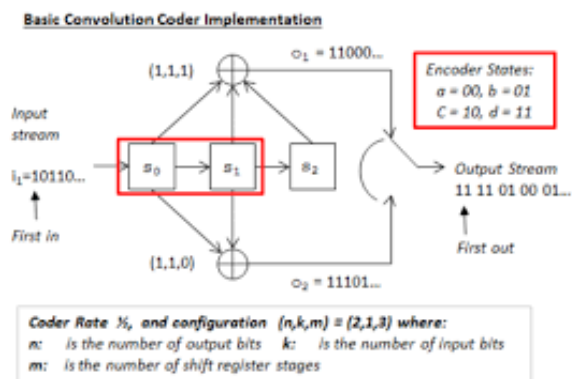

Fig.1 Basic Convolution Coder Implementation

### 4.3 Impact on Code Performance

- Error Correction Capability: Larger constraint lengths allow for better performance in correcting both random and burst errors, especially in noisy channels.

- Decoding Complexity: The complexity of the Viterbi algorithm increases exponentially with the constraint length. Thus, selecting an optimal constraint length depends on the trade-off between desired error correction and processing capability.

Design Techniques for Varying Constraint Length Codes

**Time-Varying Convolutional Codes:**

Interleavers: By permuting the data before encoding, time-varying convolutional codes can effectively spread the effects of channel bursts, improving performance in burst-error channels.

Cyclic Shifting: Cyclically shifting the encoder's state at different time instants can introduce additional randomness and enhance the code's error-correcting capabilities.

**Multi-Rate Convolutional Codes:**

Rate Adaptation: Multi-rate codes can adapt their information rate to match the channel conditions, allowing for efficient use of resources.

Code Puncturing: By deleting selected bits from the encoded sequence, the code rate can be increased to match the channel capacity.

**Turbo Codes with Varying Constraint Lengths:**

Component Encoder Selection: Turbo codes can employ component encoders with different constraint lengths to balance performance and complexity.

Interleaver Design: The interleaver can be designed to optimize the performance of the turbo code under varying channel conditions.

**Performance Analysis and Optimization**

Bit Error Rate (BER) Performance: The BER of varying constraint length codes can be evaluated using simulations or analytical techniques. Decoding Complexity: The complexity of decoding algorithms, such as Viterbi decoding, can be analyzed to assess the computational overhead associated with varying constraint lengths. Optimization Techniques: Optimization algorithms, such as genetic algorithms or simulated annealing, can be used to find the optimal parameters for varying constraint length codes, considering factors like BER, decoding complexity, and resource constraints.

## 5. Decoding of Convolutional Codes

The Viterbi algorithm is the most commonly used method for decoding convolutional codes. Its efficiency is highly dependent on the constraint length. Other algorithms, such as the BCJR (Bahl-Cocke-Jelinek-Raviv) algorithm, may be used for maximum a posteriori (MAP) decoding, but they also scale in complexity with larger constraint lengths.

## 6. Simulation and Results

A series of simulations were conducted to analyze the performance of convolutional codes with varying constraint lengths. The parameters tested include bit-error rate (BER) performance under different signal-to-noise ratios (SNRs) for constraint lengths ranging from 3 to 9. The results show that:

Codes with larger constraint lengths achieve better BER performance at lower SNRs. The computational complexity and memory requirements increase significantly for codes with constraint lengths greater than 7, which may not be suitable for real-time applications.

## Trade-offs and Design Considerations

The choice of constraint length in convolutional code design involves balancing:

Performance vs. Complexity: A longer constraint length provides stronger error correction but increases the complexity of the encoder and decoder.

Latency: Longer constraint lengths introduce additional delay in the system due to the extended memory requirements.

Application Requirements: Depending on the communication system (e.g., satellite communication, mobile networks, IoT), different constraint lengths may be preferable.

## 8. Case Study: Convolutional Codes in Satellite Communication

Satellite communication plays a vital role in global telecommunication systems, offering long-range data transmission. A key challenge in satellite communication is the presence of noise and interference that can distort transmitted signals. Error detection and correction mechanisms become crucial in maintaining signal integrity. One such powerful error-correction technique is the use of convolutional codes.

Convolutional coding is particularly effective in noisy environments like satellite communication, where signals traverse through long distances and are subject to various disturbances. These codes help in correcting transmission errors without requiring retransmission, thus ensuring reliable communication.

**Convolutional Codes: Overview**

Convolutional codes are a type of error-correcting code that encodes data by combining bits from the input sequence to produce a series of encoded output bits. Unlike block codes, where data is divided into blocks and coded separately, convolutional codes process continuous data streams. The encoded output is generated based on both the current input bit and the previous bits, depending on the constraint length of the encoder.

A convolutional encoder can be represented using shift registers and modulo-2 adders (XOR gates). These adders help combine input and stored bits to produce multiple outputs for each input bit.

**Key Parameters:**

Constraint Length (k): The number of input bits that influence the current output.

Code Rate (r): The ratio of input bits to the number of encoded output bits.

Generator Polynomial: Defines how the input bits are combined in the encoder to produce output bits.

Trellis Diagram: A graphical representation used to decode the convolutional codes.

**Application of Convolutional Codes in Satellite Communication**

Convolutional codes are widely used in satellite communication due to their ability to correct errors introduced by noise, interference, and signal attenuation in the space environment. In satellites, the long distance between the transmitter and receiver makes the signal vulnerable to fading, thermal noise, and other distortions. Convolutional codes combined with Viterbi decoding offer robust protection against such errors.

In a satellite communication system, the data (voice, video, or telemetry) is encoded using convolutional codes before transmission. The encoded data is transmitted to the ground station, where the received data is decoded using the Viterbi algorithm, which estimates the most likely transmitted sequence based on the received signal.

**Encoder Design**

In a typical convolutional encoder used in satellite systems, each input bit influences multiple output bits. The convolutional encoder consists of several shift registers and modulo-2 adders, as shown in the diagram below.

Convolutional Encoder

Here is a diagram showing the structure of a simple convolutional encoder with a constraint length of 3 and code rate 1/2.

*Input Bits:    1 0 1 1 0*

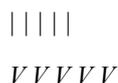

*| | | | |*

*V V V V V*

*Shift Register: [ 1 ] [ 0 ] [ 1 ]  -> [ Output Bits ]*

　　　　　*[ XOR ]      (Encoded Output)*

　　　　　*[ XOR ]*

- Shift Register: Stores previous input bits.
- Modulo-2 Adders (XOR Gates): Combine the bits stored in the shift registers to produce two encoded bits for each input bit.
- Code Rate: Since 2 bits are generated for every 1 input bit, this is a 1/2 code rate encoder.

**Trellis Diagram**

Once the signal is transmitted, convolutional codes are decoded using a trellis diagram, which represents the possible states of the encoder at different time intervals. The trellis diagram helps the decoder trace back the most likely sequence of states (and hence the input bits) based on the received signal.

**Trellis Structure**

In the trellis diagram, each state transition is associated with a set of encoded output bits. The Viterbi algorithm follows the shortest path through this trellis to decode the received signal.

*State A (00) -----> State B (01)*

　　|　　　　　|

　　v　　　　　v

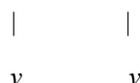

*State C (10) -----> State D (11)*

- States: Represent the contents of the shift registers.
- Transitions: Correspond to the input bits and the resulting encoded output.

Viterbi Decoding in Satellite Communication

The Viterbi algorithm is a maximum likelihood decoding algorithm used to decode convolutional codes. It traces through the trellis diagram, finding the path that minimizes the error between the received and expected signal sequences. The algorithm works by calculating a metric at each trellis stage, selecting the path with the lowest cumulative metric.

In satellite communication, the Viterbi decoder helps mitigate the effects of noise and interference by correcting transmission errors, enabling reliable data transmission even under harsh environmental conditions.

**Advantages of Using Convolutional Codes in Satellite Communication**

Error Correction: Convolutional codes significantly improve the error-correction capability, allowing satellites to communicate reliably even in the presence of high noise levels.

Efficiency: Convolutional codes offer a good balance between computational complexity and error-correction performance, making them suitable for real-time applications.

Low Latency: As convolutional codes are continuous codes, they do not require retransmission of the entire message in case of errors, reducing latency.

Adaptability: These codes are adaptable to varying channel conditions, making them effective for both

low- and high-noise environments typical in satellite communications.

Convolutional codes play a critical role in ensuring reliable communication in satellite systems. By offering strong error-correction capabilities and combining them with efficient decoding techniques like the Viterbi algorithm, they help overcome the challenges posed by noise, signal attenuation, and interference in space. As satellite communication continues to evolve, convolutional codes will remain a fundamental tool for achieving robust and efficient data transmission.

This paper presents a comprehensive analysis of convolutional code design with varying constraint lengths. The simulation results demonstrate that while larger constraint lengths offer superior error correction, they impose significant complexity on the decoding process. For practical applications, the optimal constraint length depends on the specific system requirements, including noise tolerance, available processing power, and latency constraints. Future work may explore hybrid coding schemes and machine learning techniques to dynamically adjust the constraint length based on channel conditions.

## 9. Conclusion